\documentclass[epj]{svjour}
\usepackage{graphicx}
\usepackage{amsmath}
\newcommand{\be}{\begin{equation}}
\newcommand{\ee}{\end{equation}}
\newcommand{\ben}{\begin{eqnarray}}
\newcommand{\een}{\end{eqnarray}}

\newcommand{\nd}{\noindent}
\begin{document}

\title{ Unravelling the size distribution of social groups with
information theory on complex networks}

\author{A. Hernando\inst{1}\thanks{alberto@ecm.ub.es}, D. Villuendas\inst{2}, C. Vesperinas\inst{3}, M. Abad\inst{1} \and A. Plastino\inst{4}\thanks{plastino@fisica.unlp.edu.ar}}

\institute{Departament ECM, Facultat de F\'{\i}sica,
Universitat de Barcelona. Diagonal 647,
08028 Barcelona, Spain
\and
Departament FFN, Facultat de F\'isica, Universitat de Barcelona,
Diagonal 647, 08028 Barcelona, Spain
\and
Sogeti Espa\~na, WTCAP 2, Pla\c ca de la Pau s/n, 08940
Cornell\`a, Spain
\and
National University La Plata, IFCP-CCT-CONICET, C.C.
727, 1900 La Plata, Argentina
}

\abstract{ The minimization of Fisher's information (MFI) approach
of Frieden {\it et al}. [Phys. Rev. E {\bf 60} 48 (1999)] is applied
to the study of  size distributions in social groups on the basis of
a recently established analogy between  scale invariant systems and
classical gases [arXiv:0908.0504]. Going beyond the ideal gas
scenario is seen to be tantamount to simulating the interactions
taking place in a network's competitive cluster growth process. We
find a scaling rule that allows to classify the final cluster-size
distributions using only one parameter that we call the
\emph{competitiveness}. Empirical city-size distributions and
electoral results can be thus reproduced and classified according to
this competitiveness, which also allows to correctly predict
well-established assessments such as the ``six-degrees of
separation'', which is shown here to be a direct consequence of the
maximum number of stable social relationships that one person can
maintain, known as Dunbar's number. Finally, we show that scaled
city-size distributions of large countries follow the same universal
distribution. }


\PACS{
{89.70.Cf}{Entropy and other measures of information}\and
{05.90.+m}{Other topics in statistical physics, thermodynamics, and nonlinear dynamical systems}\and
{89.75.Da}{Systems obeying scaling laws}\and
{89.75.-k}{Complex systems}
}

\date{\today}

\titlerunning{Unravelling the size distribution of social groups}
\authorrunning{A. Hernando et al.}

\maketitle

\section{Introduction}

Regularities reflected in either scaling properties \cite{bat1} or
power laws \cite{city1,power,pareto}  appear in different scenarios
related to social groups. One of the most intriguing is Zipf's law
\cite{zipf}, a power law with exponent $-2$ for the density
distribution function that is observed in describing urban
agglomerations~\cite{ciudad} and firm sizes all over the
world~\cite{firms}. This fact has received a remarkable degree of
attention in the literature. The above mentioned regularities have
been detected in other contexts as well, ranging from percolation
theory and nuclear multi-fragmentation~\cite{perco} to the
abundances of genes in various organisms and tissues~\cite{furu},
the frequency of words in natural languages~\cite{zipf,zip}, the
scientific collaboration networks~\cite{cites}, the total number of
cites of physics journals~\cite{nosotrosZ}, the Internet
traffic~\cite{net1} or the Linux packages links~\cite{linux}. More
recently, R. N. Costa Filho \emph{et al.}~\cite{elec1} found another
special regularity in the density distribution function of the
number of votes in the Brazilian elections, a power law with
exponent $-1$. This law has also been found in Ref.~\cite{nosotros},
using an information-theoretic methodology \cite{alp}, for both the
city-size distribution of the province of Huelva (Spain) and the
results of the 2008 Spanish General Elections. These findings allow
one to conjecture that this behavior reflects
 a second class of universality.

\nd What all these disparate systems have in common is the lack of
a characteristic size, length or frequency for the observable
under scrutiny, which makes them scale-invariant. In
Ref.~\cite{nosotros} we have introduced an information-theoretic
technique based upon the minimization of Fisher's information measure
\cite{alp} (abbreviated as MFI) that allows for the formulation of
a ``thermodynamics" for scale-invariant systems. The methodology
 establishes an analogy between such systems  and
physical gases which, in turn,  shows that the two special power
laws mentioned in the preceding paragraph lead to a set of
relationships formally identical to those pertaining to the
equilibrium states of a scale invariant non-interacting system,
the \emph{scale-free ideal gas} (SFIG). The difference between the
two  distributions is thereby attributed  to different boundary
conditions on the SFIG.

\nd However, there are many social systems that can not be
included into any of these two universality classes and exhibit
different kinds of behavior~\cite{nosotros}. In order to deal with
 them,  during the last years researchers have worked out  different
mathematical models and thus  addressed  urban dynamics~\cite{ud}
and electoral results~\cite{elec1,elec2}, developing detailed
realistic approaches.  Ref.~\cite{otros} is highly recommendable
as a primer on urban modelling. However, some aspects of the
concomitant problems defy full understanding, since a clear
prescription for the classification of the size-distribution of
social groups is  still missing. To remedy such an understanding-gap
is our main purpose here.

\nd The goals and motivation of this work are thus focussed  on
gaining insight into such size-distributions in the case of systems
that can not be described by recourse to the two power laws
described above, i.e., by a {\it non-interacting} scenario. If an
analogy with real gases is worked out  when interactions are duly
taken into account, a microscopic description is needed in order to
obtain the pair correlation function~\cite{MD}. This is achieved
using numerical simulations as in molecular dynamics. An similar
path will be followed here by recourse to the Fisher-derived analogy
of~\cite{nosotros}. Thereby we go beyond the SFIG stage by using a
proportional growth process (PGP) so as to model the interaction
between the elements of the social network system. We can thus study
the PGP effect on the density distribution. This requires to have at
hand a way to properly describe scale invariance at the microscopic
level~\cite{nosotros,exp} via a  competitive cluster growth process
within a complex network.

\nd This work is organized as follows: in Sec. II we describe the
application of the MFI approach \cite{alp} to complex networks in
order to obtain the degree distribution  and thus describe the
competitive cluster growth process (inside the network). This allows
one to, in turn,
 microscopically simulate growth processes in a social group. In
Sec. III we study the size distributions obtained using this
methodology. We find a scale transformation that allows for
systematically classifying the deviations from the SFIG that we
encounter  in the cluster-size distributions. This classification is
effected using just a single parameter, which we call the
\emph{competitiveness}. We also apply this criterion to classify the
city-size distributions of the provinces of Spain and some electoral
results. Moreover, we show that empirical assessments as the average
path length and Dunbar's number are well reproduced by our approach.
Using such a scale transformation we demonstrate  that most
distributions of city population in large countries exhibit the same
shape. Finally, in Sec.~IV we  draw some  conclusions.

\section{Theoretical method}

\nd City-size distributions and electoral results display a
similar scale-free behavior, and both of them have the same
constituents: groups of people. Although the resources of these
groups or the interests of the individuals composing them may be
different in each case, a naive approach is to assume that people
are connected to other people, hence giving rise to a network
where groups of interest develop. Network theory \cite{net1} has
been successfully used before for dealing with electoral results
and the spread of opinions, which encourages to employ it to develop
the microscopic description of the associated systems.

\subsection{The scale free ideal network}

\nd The basic elements of networks are ``nodes" that are connected
to other nodes by ``edges". The degree $k$ of each node is defined
as the number of connections  it possesses. The degree distribution
(DD) $F(k)$ and the way the nodes are connected define the
statistical properties of the network. Scale-free complex networks
display many interesting properties that have been found in
techno-sociological systems such as the Internet (World Wide Web
\cite{bbaa}, e-mail networks \cite{email} and also
instant-message-sending networks \cite{mess}, for example).

\nd In our model, we assume that the network can be described at
the macroscopic level as a scale invariant system of $N$ nodes,
with the number of connections $k$ the ``coordinate" that locates
each node in the pertinent configuration space. We also assume
that the degree of each node does not depend on the degree of
other nodes. In these circumstances we can i) legitimately describe
the network as a SFIG in an equilibrium state, a scenario to be
denoted as the \emph{scale free ideal network} (SFIN), and ii)
derive the DD via the MFI, which we pass now to recapitulate.

\subsubsection{Minimum Fisher Information approach (MFI)}

\nd The Fisher information measure $I$ for a system of $N$
elements, described by the coordinate $k$ and the physical
parameters $\theta$ has the form~\cite{libro}
\begin{equation}\label{fish}
I(F)=c_k\int dkF(k|\theta)\left|\frac{\partial\ln
F(k|\theta)}{\partial \theta}\right|^2,
\end{equation}
where $F(k|\theta)$ is the density distribution in configuration
space and the constant $c_k$ accounts for proper dimensionality.
According to MFI tenets \cite{alp}, the equilibrium state of the
system minimizes $I$ subject to prior conditions, such as the
normalization of $F$, namely $\langle 1 \rangle=1$. The MFI is then
cast as a variation problem of the form \cite{alp}
\begin{equation}
\delta\left\{I(F)-\mu\langle 1 \rangle\right\}=0,
\end{equation}
where $\mu$ is the normalization-associated Lagrange multiplier.

\subsubsection{Application of the MFI to a scale-free network}

In the case of the derivation of the DD of our complex network, we
define a minimum degree of unity and a maximum degree value of
$k_M$. With the change of variable $u=\ln k$, the scale
transformation $k'=k/\theta_k$ transforms $u$ into
$u'=u-\Theta_k$, where $\Theta_k=\ln\theta_k$. The distribution of
physical elements is then described by the mono-parametric
translation families $F(k|\theta)=f(u|\Theta_k)=f(u')$. Taking
into account the fact  that the Jacobian of the transformation is
$dk=e^udu$, the information measure $I$ can be obtained in the
continuous limit as
\begin{equation}
I=c_u\int_0^{\ln k_M}\!\!\!du~e^{u}f(u)\left|\frac{\partial\ln
f(u)}{\partial u}\right|^2,
\end{equation}
and, with the normalization constraint
\begin{equation}\label{w2a}
\int_0^{\ln k_M}\!\!\!du~e^{u}f(u)=1,
\end{equation}
the variation problem reads now
\begin{equation}
\delta\left\{c_u \int_0^{\ln k_M}\!\!\!du~e^{u}f\left|\frac{\partial\ln
f}{\partial u}\right|^2+\mu \int_0^{\ln k_M}\!\!\!du~e^{u}f\right\}=0.
\end{equation}
Introducing $f(u)=e^{-u}\Psi^2(u)$, and varying with respect to
$\Psi$ leads to the Schr\"odinger-like equation \cite{alp}
\begin{equation}
\left[-4\frac{\partial^2}{\partial u^2}+1+\mu'\right]\Psi(u)=0,
\end{equation}
where $\mu'=\mu/c_u$. The general solution to this equation is
$\Psi(u)=e^{-\alpha u/2}$ with $\alpha=\sqrt{1+\mu'}$. Equilibrium
corresponds to the ground state solution $\alpha=0$ \cite{alp},
which yields the same density distribution as that of the SFIG in
the thermodynamic limit
\begin{equation}
F(k)dk=\frac{1}{\ln k_M}\frac{dk}{k}.
\end{equation}

\nd Once we know, for a given total number of nodes $N$, the
associated DD of the SFIN, we proceed  by assigning a number of
potential edges $k$ to each node, with $k$ randomly obtained from
$F(k)$. Accordingly,  the $N$ nodes are randomly connected among
themselves by their assigned edges, with two restrictions: a node
cannot be connected to itself nor twice to the same node. The
ensuing process ends when no more connections can be established.
We will show later that the values of $N$ and $k_M$ can be
arbitrarily chosen in order to classify the empirical
distributions.

\begin{figure}[tbp]
\center
\includegraphics[width=\linewidth,clip=true]{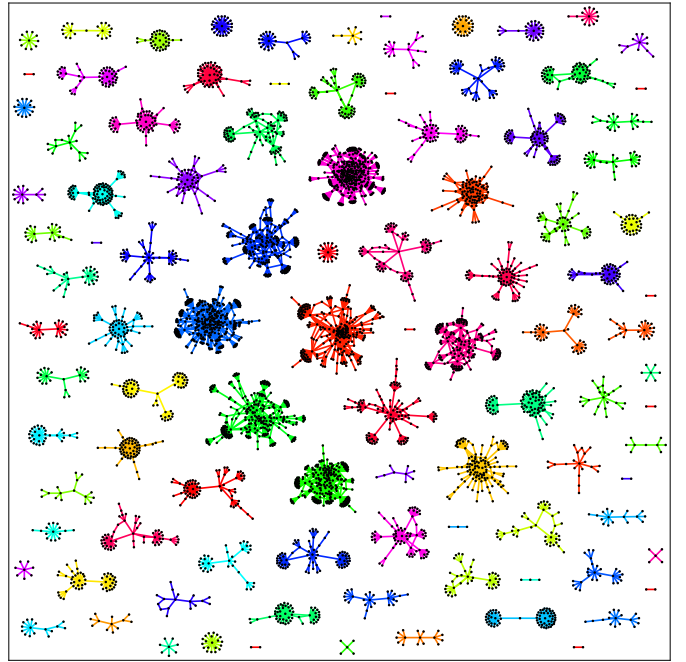}
\caption{(Color online) Final result of a competitive process of
$n_c=100$ clusters with an initial size of $n_i=1$ nodes in a
network of $N=5\,000$ nodes. All the clusters belong to the same
network and are connected among themselves. For clarity's sake,
however, they have been plotted independently. A large variety of
cluster sizes ensues.\label{fig1}}
\end{figure}

\subsection{The competitive cluster growth process}

\nd Once we have built a SFIN with $N$ nodes and maximum degree
$k_M$, we apply a competitive cluster growth process to it, as
discussed in~\cite{bran}. This technique falls into the category of
PGP or \emph{discrete multiplicative processes}, which are known to
correctly describe scale-invariant behavior~\cite{exp}. For
starters, we fix the values of the minimum cluster size $n_i$ and
the total number of clusters $n_c$ that will grow in the network.
Next, $n_c\times n_i$ nodes of the network are randomly selected  as
cluster ``seeds". In the first iteration the first neighbors of the
seeds are incorporated to the cluster in random order, unless they
are seeds of other clusters. At subsequent iterations, the first
neighbors of the nodes added at the precedent step are, in turn,
randomly added to the cluster, unless they already belong to a
different cluster. The process ends when all the nodes belong to
some cluster. We display in Fig.~\ref{fig1} the final result of a
competitive growth process for $n_c=100$ clusters with an initial
size of $n_i=1$ node in a network of $5\,000$ nodes.

\nd The procedure  may include a probability $t$ for  a node
changing to another cluster if any of its neighbors belongs to
this cluster (micro-dynamics). However, M. Batty found that even
if the rank of the cities rapidly evolves  in time due to
micro-dynamics, the city-size distribution evolves only slowly
\cite{clock}: the system evolves quasi-statically at the
macroscopic level. We then consider that the system exhibits an
adiabatic evolution, implying that our distributions can be well
represented  by stationary configurations ($t=0$). Although  not
at this stage,  we expect to study micro-dynamics  in the
future.\\

\section{Present results}

\subsection{Study and classification of cluster-size distributions}

\subsubsection{Regime of low-density of clusters: recovering the SFIG}

\nd We have studied the size distribution of SFIN-clusters with
$N$ and $k_M$ ranging from $N=5\,000$ to $500\,000$ and $k_M=50$ to
$500$ ---finite-size effects may be relevant for smaller values of
$N$ and $k_M$. When the density of seeds,
defined as $\rho_s=n_c n_i/N$, is much lower than unity, the
probability density distribution of sizes $p(x)$ mostly follows that of the
SFIG at equilibrium, which is in the continuous limit
\begin{equation}\label{distri}
p(x)dx =
\begin{cases}
\dfrac{1}{\Omega}\dfrac{dx}{x}  &\text{if } x_1 \leq x \leq x_M\\
0                        &\text{otherwise}
\end{cases},
\end{equation}
where $\Omega=\ln(x_M/x_1)$ is the ``volume" in the concomitant
size-configuration space. The maximum  $x_M$ and minimum $x_1$
sizes generally depend on $n_i$, $n_c$, and $N$. Since finite-size
effects make it difficult to estimate $x_1$ and $x_M$, we have found it
useful to evaluate the volume as $\Omega=2\ln(x_{3/4}/x_{1/4})$,
where $x_{1/4}$ and $x_{3/4}$ indicate the first and third
quartiles of the distribution.

\nd It is convenient for a scale-invariant system
to introduce a new variable $x'=x/\theta$ without changing
the physics, with $\theta$ a parameter to be later defined.
Furthermore, we can rescale the volume to $\Omega'=C\Omega$
according to
\begin{equation}\label{trans1}
x' =\left(\frac{x}{\theta}\right)^C,
\end{equation}
which leads to the scaled distribution
\begin{equation}
p(x')dx' =
\begin{cases}
\dfrac{1}{\Omega'}\dfrac{dx'}{x'}  &\text{if } \left(\frac{x_1}{\theta}\right)^C \leq x' \leq\left(\frac{x_M}{\theta}\right)^C\\
0                        &\text{otherwise}
\end{cases}.
\end{equation}
Note that these changes do not affect the properties of the
distribution, which remains that of a SFIG. It is also useful
 to employ the reduced units defined by $\theta=x_{1/2}$ and
$\Omega'=2$, where $x_{1/2}$ is the median of the distribution. In
this particular case, for the new variable $y$ defined by the
transformation
\begin{equation}\label{trans0}
y=\left(\frac{x}{x_{1/2}}\right)^{\frac{1}{\ln\left(\frac{x_{3/4}}{x_{1/4}}\right)}},
\end{equation}
the density distribution takes the form
\begin{equation}
p(y)dy =
\begin{cases}
\dfrac{1}{2}\dfrac{dy}{y}  &\text{if } e^{-1} \leq y \leq e\\
0                        &\text{otherwise}
\end{cases}.\label{Distpy}
\end{equation}
For convenience we define a ``normalized" rank-parameter $r$ in such
a way that all the pertinent ``sizes" to be here considered range
within the interval [0,1]. This normalized rank-size distribution
associated to the density distribution gets cast as
\begin{equation}
y = e^{1-2r}.\label{Disty}
\end{equation}
Note that the density distributions \eqref{Distpy} and associated
the rank-size \eqref{Disty} do no longer depend on $n_i$, $n_c$,
$N$, or $k_M$, since $x_M$ and $x_1$ do not enter  the definition of
the maximum and minimum sizes when expressed in such units.

\subsubsection{Regime of high density of clusters: classification by competitiveness}

\nd When we increase the number of clusters, the competition
for space grows and the size of a cluster depends now on the size
of the neighboring clusters. The size distribution exhibits
important deviations from the SFIG, but the change to reduced
units makes it still possible to compare between distributions
obtained with different values of $n_i$, $n_c$, $N$ and $k_M$.
These comparisons have led us to find a classification of the
distributions using a parameter $\lambda$ ---which we denominate
\emph{competitiveness}--- that we pass now to discuss.

\begin{figure}[tbp]
\center
\includegraphics[width=\linewidth,clip=true]{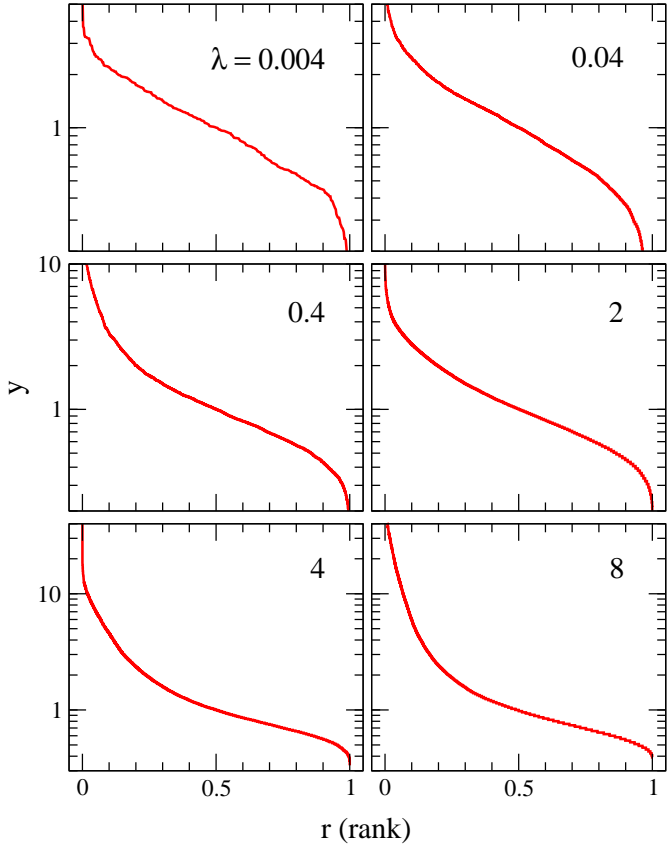}
\caption{(Color online) Scaled rank-size distribution
$y(\lambda,r)$ in semi-log scale for different values of
competitiveness $\lambda$. These curves have been obtained by
computing a large number of competitive processes to reduce
statistic fluctuations.\label{fig2} }
\end{figure}

\nd Network configuration theory tells us that for a
given degree distribution, the mean number of $j$-th neighbors
of a node is~\cite{net1}
\begin{equation}\label{eq:estate}
z_j=\left(\frac{z_2}{z_1}\right)^{j-1}z_1,
\end{equation}
where $z_1$ and $z_2$ are the mean number of first and second
neighbors, respectively. Consequently, the mean size $\langle x \rangle_s$ of the cluster
generated for each seed is, at the end of the process,
\begin{equation}\label{eq:estate2}
\langle x \rangle_s=\sum_{j=1}^{j_f}z_j=\left[\sum_{j=1}^{j_f}\left(\frac{z_2}{z_1}\right)^{j-1}\right]z_1=\lambda^{-1}z_1
\end{equation}
where $j_f$ is the mean number of total iterations used in the
process, and $\lambda^{-1}$ is a new parameter defined by
(\ref{eq:estate2}) for future convenience. Since all nodes of the
network belong, at the end of the process, to a certain cluster, the
mean size times the number of seeds must be equal to the total
number of nodes, i.e.,
\begin{equation}\label{eq:estate3}
n_cn_i\lambda^{-1}z_1=N.
\end{equation}
For a scale-free ideal network, $z_1=\langle k\rangle=(k_M-1)/\ln
k_M$, which gives for large $k_M$
\begin{equation}\label{eq:estate4}
\lambda = \frac{n_c n_i}{N} \frac{k_M}{\ln k_M}=\rho_s\frac{k_M}{\ln k_M}.
\end{equation}
We interpret $\lambda$ as a quantifier of the strength of the
interactions and use it to classify the family of distributions
obtained via our simulations.
In our simulations we have studied distributions with
values ranging from $\lambda\rightarrow0$
---where the SFIG emerges naturally--- up to $\lambda\sim 10$
for very high density and a very connected network ---or very
small-world  \cite{small}. Anyhow, we have found no evidence of an
upper bound in $\lambda$. We display in Fig.~\ref{fig2} the
rank-size $y(\lambda,r)$ in semi-log scale for different values of
competitiveness  $\lambda$. These curves have been obtained by
generating several networks and computing a large number of
competitive processes within them to reduce numerical
fluctuations.

\subsection{Study and classification of empirical distributions by competitiveness}

\nd We have found that the change performed above to reduced unit
$y$ can be applied to empirical data to compare the distributions
for city sizes and electoral results in different countries or
states. It allows to compare the effect of societal features, such
as policies and economic data. Furthermore, a comparison between
these distributions and those obtained via our simulations can be
performed to assign a competitiveness value to the empirical data.
This assignation is effected by minimizing the distance between the
data and the computed curve $y(\lambda,r)$, using a
Kolmogorov-Smirnof test~\cite{KST} (some examples are displayed in
Fig.~\ref{fig3}-Fig.~\ref{fig9} for electoral results and city
populations). Since our simulation fits nicely  the data, we are
compelled to conclude that \emph{in general, the scaled
distributions of city populations and electoral results can be
classified according to the values of  $\lambda$}.

\subsubsection{City size distributions}

\begin{figure}[tbp]
\center
\includegraphics[width=\linewidth,clip=true]{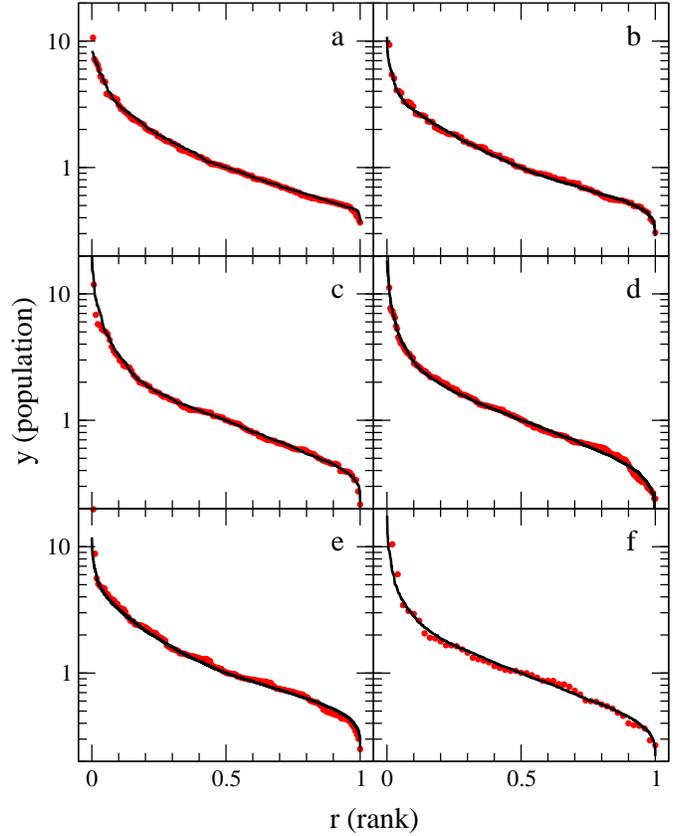}
\caption{(Color online) Classification of city-size distributions by
competitiveness ($\lambda$). \textbf{a}, scaled rank-plot of
the city population of Girona province ($\lambda=1.74$)
\textbf{b}, Bizkaia ($\lambda=1.63$) \textbf{c}, Castell\'o
($\lambda=0.65$) \textbf{d}, Cuenca ($\lambda=0.58$) \textbf{e},
Granada ($\lambda=0.06$) \textbf{f}, capital cities of  Spanish
provinces ($\lambda=0.71$). All  empirical data are plotted with red
dots, and compared to the rank-size distributions obtained with a
numerical simulation employing  the same value of competitiveness
(in black lines).\label{fig3} }
\end{figure}

\begin{figure}[tbp!]
\center
\includegraphics[width=\linewidth,clip=true]{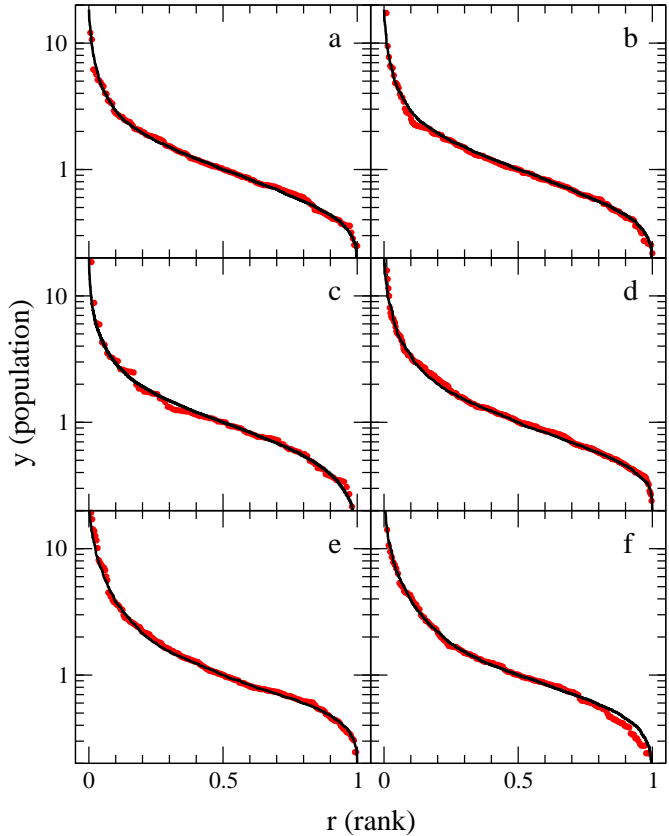}
\caption{(Color online) Classification of city-size distributions by
competitiveness ($\lambda$). \textbf{a}, scaled rank-plot of
the city population of Tarragona province ($\lambda=0.69$)
\textbf{b}, C\'aceres ($\lambda=0.65$) \textbf{c}, Burgos
($\lambda=0.65$) \textbf{d}, Cantabria ($\lambda=0.59$) \textbf{e},
\'Avila ($\lambda=0.43$) \textbf{f}, La Rioja ($\lambda=0.33$). All
empirical data are plotted using red dots, and  compared to the
rank-size distributions obtained with our numerical simulation
employing  the same value of competitiveness (in black
lines).\label{fig4} }
\end{figure}

\nd We have performed an exhaustive city-population study for the
provinces of Spain \cite{muni}. We have fitted each distribution to
$y(\lambda,r)$ and have found a competitiveness distribution with a
median of $\lambda_{1/2}=0.65$, reflecting some local dependence. We
depict in Figs.~\ref{fig3} and~\ref{fig4} the scaled rank-size
distributions of some provinces, together with the accompanying
$\lambda$-family of distributions, which nicely fit the data. We
have found that the rank-size distribution of the capital cities has
a competitiveness of $0.71$ (Fig.~\ref{fig3}f), which does not
significantly differ from the median value. We contend that the fact
that this distribution can be classified by competitiveness
is a signature of the scale invariant nature of the social system:
the whole country can be thought of as a single network of (only)
capital cities, which displays similar statistical properties as
those of the complete network, which includes all cities.

\begin{figure}[tbp!]
\center
\includegraphics[width=\linewidth,clip=true]{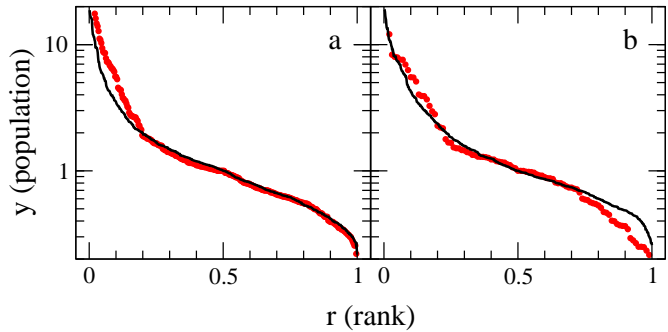}
\caption{(Color online) Examples of deviations from the $\lambda$-
family of distributions. \textbf{a}, scaled rank-plot of the city
population of Guadalajara province, compared with the best fit.
Deviations are seen in the case of the cities at the graph's top
(see text).
 \textbf{b}, M\'alaga, where the above deviations are also present. All
 empirical data are plotted using red dots, and simulation with
black lines.\label{fig5} }
\end{figure}

\nd We have detected  some singular exceptions in the fitting of
these curves, as illustrated in Fig.~\ref{fig5} for the provinces of
Guadalajara and M\'alaga. We understand that these deviations from
the $\lambda$-family of distributions reflect local effects in
policies or in social, economical or geographical factors, as some
studies have found~\cite{valitova}. In the case of Guadalajara, the
uppermost cities in the plot ---those that deviate from the best
fit--- are located in the neighborhood of Madrid, the capital of
Spain. The capital of Spain thus affects the population distribution
in its neighborhood.

\begin{figure}[tbp]
\center
\includegraphics[width=0.8\linewidth,clip=true]{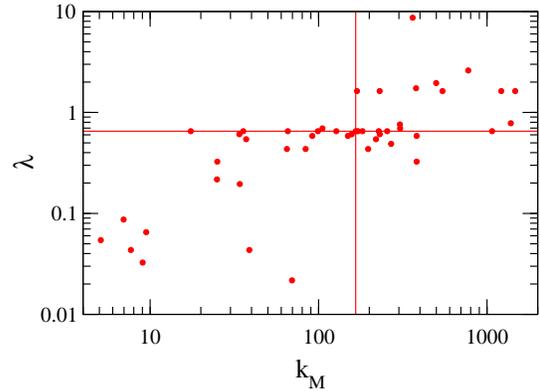}
\caption{(Color online) Plot of the competitiveness $\lambda$ of the
Spanish provinces versus the empirical value of maximum degree
$k_M$. The medians of both parameters are represented by
lines.\label{fig6} }
\end{figure}

\begin{figure}[tbp]
\center
\includegraphics[width=0.8\linewidth,clip=true]{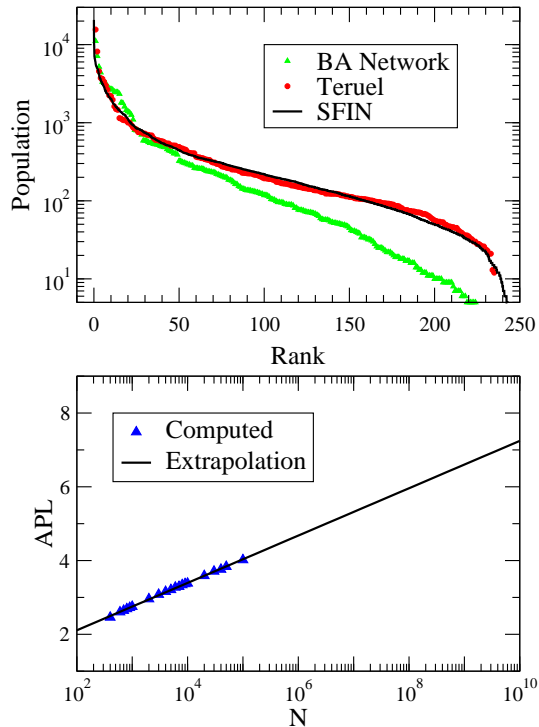}
\caption{(Color online) Top panel, rank-size distribution of Teruel,
Spain (red dots) is compared to the result of competitive processes
carried out in networks with $N=100\,000$ nodes and where Dunbar's
number is the maximum degree, $k_M=150$ (black line). The
size distributions are not scaled, which implies a relation one to
one between inhabitants and nodes. We also show the result of the process
in a BA network, which is not able to reproduce the empirical
distribution (see text). Bottom panel, average
path length ---computed (blue dots) and extrapolated (black line)--- for
networks with maximum degree $k_M=150$. The extrapolation (see text)
reproduces recent empirical measures.\label{fig7} }
\end{figure}

\nd  The empirical maximum degree $\hat{k}_M$, which defines the
volume of the SFIN in configuration space, has been estimated for
each province by solving the equation
\begin{equation}\label{km}
 \frac{\hat{k}_M}{\ln\hat{k}_M}  = \lambda\frac{\hat{N}}{\hat{n}_c\hat{n}_i}.
\end{equation}
Here $\hat{N}$ is the total population, $\hat{n}_c$ the number of
cities and $\hat{n}_i$ the population of the smallest city, which is
used to estimate the minimum cluster size. We have found that the
empirical distribution of the maximum degree exhibits a large tail,
whose median is $166$. The first and third quartiles are $37$ and $319$
respectively, hence
\begin{equation}
\langle\hat k_m\rangle = 166^{+153}_{-129}.
\end{equation}
We exhibit in Fig.~\ref{fig6} the competitiveness versus the
empirical value of maximum degree of the Spanish provinces. The
medians of both parameters are also shown.
The maximum number of connections is an observable that has been
evaluated before in the literature and is known as Dunbar's
number~\cite{resdum}. It can be found in the fields of
anthropology, evolutionary psychology, and sociology. It
reflects the fact that the maximum number of individuals with whom
any person can maintain stable social relationships is determined
by the size of their neocortex \cite{dunbar}. Dunbar's number
lies between $100$ and $230$, but a commonly detected  value is
$150$, which fits quite well our results. As far as we know,
\emph{the present work is the first in which Dunbar's number is
computed using a mathematical model based on first principles.} We
have checked with the case of the province of Teruel that a SFIN
with a maximum degree of $150$ is able to reproduce the associated
empirical distribution without the need of scaling it via the
change to variable $y$. A total population of $109\,810$
inhabitants, excluding the capital city --we have already seen
that the capital belongs to a larger national network--,
distributed into $235$ cities, is modelled by a network of i)
$N=100\,000$ nodes, ii) a maximum degree of $k_M=150$, and iii) by
``growing" $n_c=250$ clusters with an initial size of $n_i=1$
node. We depict the rank size distribution in Fig.~\ref{fig7}a,
where it can be clearly seen that the simulation nicely fits the
data. The city-size distribution is also compared with the cluster
growth process obtained in a Barabasi-Albert network (BA) with the same
number of nodes and clusters. In this figure we see that
not all kinds of networks will be able
to reproduce the empirical distribution, even if we employ a similar
number of nodes and clusters, as in the case of the BA network.

\nd The average path length (APL) of a network is defined, for all
possible pairs of nodes, as the average number of steps along the
shortest path. It is one of the most important quantities
characterizing a network's topology~\cite{net1}. We have numerically
computed the APL of a SFIN with $k_M=150$ as a function of $N$ up to
$N=100\,000$ nodes. One easily sees  the expected dependence on
$\log N$, as illustrated by Fig.~\ref{fig7}b. The extrapolation
gives APL~$=4.00$ for a SFIN of $100\,000$ nodes, APL~$=5.63$ for
$45\,000\,000$ nodes (population of Spain), APL~$=6.13$ for
$300\,000\,000$ nodes (population of the USA \cite{usa}), and
APL~$=6.95$ for $6\,500\,000\,000$ nodes (World population). These
values are in accordance with the empirical measure of Travers and
Milgram, known as the ``six degrees" \cite{six}, and with the more
recent results of P.S. Dodds \emph{et al.}, who found an APL between
$5$ and $7$ \cite{ele}, or J. Leskovec and E. Horvitz, who found 6.6
degrees between Messenger users \cite{mess}. These results indicate
that \emph{the ``six degrees of separation" is a direct consequence
of Dunbar's number.}

\subsubsection{Electoral results}

\begin{figure}[tbp]
\center
\includegraphics[width=\linewidth,clip=true]{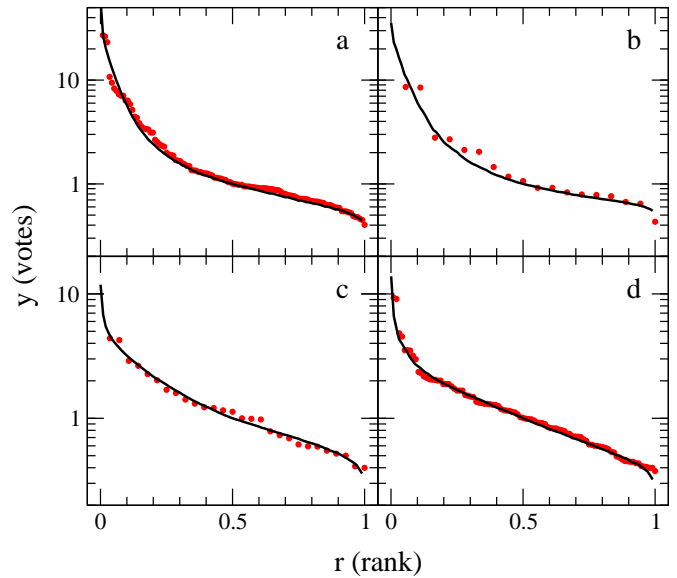}
\caption{(Color online) Classification of electoral
results distributions by competitiveness ($\lambda$).
\textbf{a}, scaled rank-plot of the 2005 elections results in UK
($\lambda=6.5$). \textbf{b}, 2004 elections results in the USA
($\lambda=4.6$). \textbf{c}, 2008 elections results in Italy
($\lambda=2.7$). \textbf{d}, 2008 elections results in Spain
($\lambda=0.98$). All  empirical data are plotted using red dots,
and are compared to the rank-size distributions obtained via a
simulation employing  the same value of competitiveness (in black
lines).\label{fig8} }
\end{figure}

\nd  We have carried out a similar competitiveness study for the
results of General Elections in different countries and computed the
$\lambda$ value in the cases of UK'05 \cite{elecUK}, USA'04
\cite{elecUS}, Italy'08 \cite{elecIt}, and Spain'08 \cite{elecSp},
finding $\lambda = 6.5$, $4.6$, $2.7$, and $0.98$, respectively
(Fig.~\ref{fig8}), all values being larger than the average found
for city populations. In general, a high value of the
competitiveness increases the difference in the number of votes
between two consecutive parties in the rank of results.

\nd  The estimates of the maximum degree are $\hat{k}_M
\sim~600\,000$, $3\,000\,000$, $19\,000$, and $35\,000$,
respectively, i.e., many orders of magnitude larger than
Dunbar's number. Thus,  the volume in configuration space of the
SFIN that describes the election process is larger than that for the
city population. This is the effect of the creation and development
of temporary connections. A politician, journalist, or blog writer
can be easily connected during the electoral campaign to thousands
of people via mass media, such as television, newspapers, or the
Internet. In accordance with Dodds' results, the world becomes
smaller ---more connected--- when individual incentives exist
\cite{ele},  in this case to obtain good electoral results. These
findings lead to interesting conclusions. In the USA's case  we  find
larger hubs than in the UK: $3\,000\,000$ connections against
$600\,000$, but since the total population is $N_p=300\,000\,000$
against $61\,000\,000$ \cite{pobuk}, the relative value is similar
for each country, $\hat{k}_M/N_p\sim0.01$. This value indicates that
the USA and the UK have similar social networks in electoral
campaigns, but scaled. Since there are more parties competing in
elections in UK's case, the distribution of the results naturally
displays a higher competitiveness than in the USA's one.

\subsubsection{The universal distribution}

\begin{figure}[tbp!]
\center
\includegraphics[width=0.8\linewidth,clip=true]{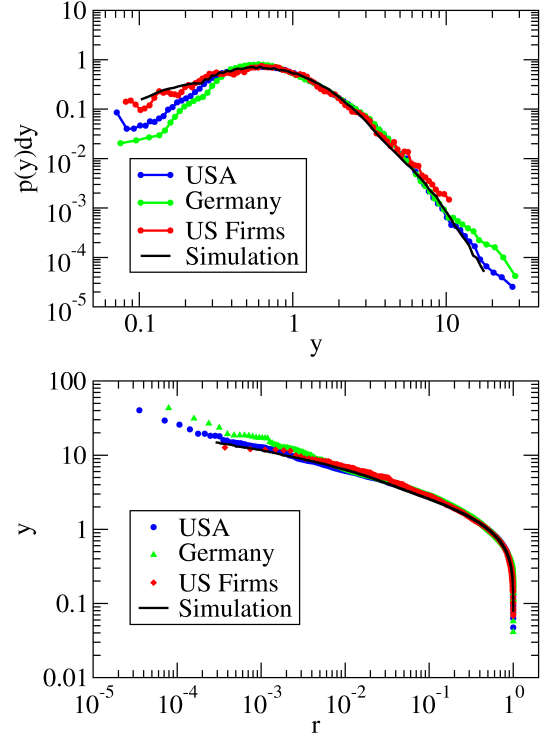}
\caption{(Color online) Top panel, scaled city-size distribution
for the USA (green squares) and Germany (blue triangles), and scaled
firm-size distribution of the USA (red circles) compared to the
universal distribution generated with our numerical simulation
(black line) by mixing a large amount of data with different
values of competitiveness. Bottom panel, same as top panel for the
scaled rank-size distribution in log-log scale.\label{fig9} }
\end{figure}

\nd  Studying the city population of different countries around the
world~\cite{datospaises} we have found that, for countries with a
population over $5\,000\,000$, the main portion of the scaled
distributions turns out to be quite similar, in fact the same
distribution, thus evidencing  some degree of universality, as
illustrated in Fig.~\ref{fig9} for  USA and Germany. Even the
distribution of the size of companies in these countries follows
this behavior, as depicted  in the same figure for USA firms
\cite{usa}. This universal distribution can be reproduced by our
simulation. Note  that the competitiveness has a local dependence,
and thus  data of a country are in fact several  sets of data (for
many states or provinces of that country), which have different
values of the competitiveness. We have simulated this universal
distribution by mixing data generated with different values of
competitiveness, between $0.4$ and $1$, obtaining the curve
$y_0(r)$, which nicely fits the empirical distributions as can be
seen in Fig.~\ref{fig9}.

\section{Summary and discussion}

\nd We have shown in this communication that the main properties of
the city-size distributions and electoral results can be well
reproduced when interactions between network elements are introduced
by means of a competitive cluster growth process in a SFIN. We
classify the deviations from the SFIN distribution in terms of just
a single parameter, the competitiveness $\lambda$, that quantifies
the strength of the interaction between the elements of the system.
As expected, the SFIG-distribution emerges naturally in the limit of
low competitiveness. The value of $\lambda$ can be easily extracted
from empirical data by using the transformation to reduced units
given in Eq.~(\ref{trans0}) and then comparing the scaled
distribution to a distribution of known competitiveness.

\nd In our simulations this parameter is related to the total
density of clusters and to the maximum degree of the network ---or
the volume in the configuration space--- by
Eq.~(\ref{eq:estate4}). For real systems, our results in the study
of the Spanish provinces indicate that this relation remains
valid. We have used it to compute the empirical average of the
maximum degree, finding that it reproduces  Dunbar's number
\cite{dunbar}. Furthermore, the rank-size distribution of Teruel
is reproduced using  real values for the  density of cities
together with a maximum degree in a SFIN. Our simulations also
predict the empirical estimate of the average path-length when we
use Dunbar's number for the maximum degree of the SFIN. This
indicates that the known ``six degrees of separation"~\cite{six} is
a consequence of Dunbar's number. For electoral results,
we have found that the maximum degree grows by an order of
magnitude
---the volume in the configuration space grows---, which confirms
the statement that the world is more connected when individual
incentives do play a role.\cite{ele}

\nd Some studies have found correlations between city-size
distribution and regional policies~\cite{valitova}. We believe
that the use of the $\lambda$ parameter for such studies would add
a very useful tool in order to classify the ensuing distributions.
What  could represent an advance in social and political sciences,
would be to  systematically assess  the dependence of the
competitiveness on local policies. As seen in the case of
electoral results, a high value of the competitiveness enhances
the difference (in number of votes) between two consecutive parties in
the results rank. This implies that a small party would prefer a
scenario with a low value of $\lambda$ in order to get better
chances in the final tallies, whereas a big party would choose  a
high value in order to increase the relative difference with the
other parties. For city sizes, a low value of the competitiveness
works against supersaturated cities, whereas a high value promotes
the importance of a capital city.

\nd In general, all empirical distributions agree quite well with
those obtained with our simulation, but  we found also some
singular exceptions. We expect these to be related to the already
mentioned regional policies, and to historical or geographical
factors. Thus, our model could help to identify such
scenarios. Exhaustive studies of data around the world are
necessary to build a bridge between the three variables of
Eq.~(\ref{eq:estate}), $\rho_s$, $k_m$ and $\lambda$, and the
social and economic polices of a region. It is also reasonable to
think that a study in competitiveness terms of the evolution of
firms-size distributions during  the last years may lead to a
deeper understanding of the present economic situation.

\nd Summing up, our results show that scale invariant
thermodynamics yields  a useful framework for dealing with  scale
invariant phenomena. Its application to social sciences here  has
provided some  deeper insight into the way humans build up a
society. This work only represents a first step, and it is
expected that subsequent studies will enhance the predictive power
of the theory.

\begin{acknowledgement}
We would like to thank Manuel Barranco for useful discussions,
and to Albert D\'iaz, Carles Panad\`es, Joan Manel Hern\'andez, Antoni Garc\'ia for their
helpful comments and remarks. This work has been partially
performed under grant FIS2008-00421/FIS from DGI, Spain (FEDER).
\end{acknowledgement}


\end{document}